\documentclass[runningheads]{llncs}
\usepackage{tikz}
\usetikzlibrary{positioning,calc}
\usepackage[table]{xcolor}
\usepackage[T1]{fontenc}
\usepackage{bm}
\usepackage{hyperref}
\usepackage{amsfonts}
\usepackage{pifont}
\usepackage{float}
\usepackage{amsmath} 
\usepackage{graphicx,verbatim}
\usepackage{xcolor}
\usepackage{cite} 
\usepackage{booktabs} 
\usepackage{booktabs}
\usepackage{pifont}
\newcommand{\cmark}{\ding{51}}
\newcommand{\xmark}{\ding{55}}
\begin{document}

\title{GazeRefine: Expert Gaze as a Test-Time Prompt for Training-Free Medical Image Segmentation}

\titlerunning{GazeRefine for Training-Free Medical Segmentation}

\author{Mohammed Oussama Benyahia$^{1}$, Marouane Tliba$^{1}$, Mohamed Amine Kerkouri$^{1}$, Taifour Yousra$^{1}$, Bin Wang$^{2}$, Max Bengtsson$^{2}$, Gorkem Durak$^{2}$, Elif Keles$^{2}$, Zuheng Ming$^{1}$, Marek Penhaker$^{3}$, Azeddine Beghdadi$^{1}$, Ulas Bagci$^{2}$, Aladine Chetouani$^{1}$} 
\authorrunning{BENYAHIA et al.}
\institute{Université Sorbonne Paris Nord$^{1}$, Northwestern University$^{2}$,  VSB-Technical University of Ostrava$^{3}$ \\
    }
  
\maketitle             

\begin{abstract}
Medical image segmentation remains difficult to scale because high-performing methods typically rely on dense expert annotations and task-specific training. We introduce \textbf{GazeRefine}, a training-free framework that uses gaze as an inference-time prompt for zero-shot medical image segmentation. Sparse, duration-weighted fixations are converted into foreground and background priors that initialize semantic prototypes in frozen DINOv3 feature space. These prototypes are iteratively refined through foreground--background discrimination, feature-space affinity propagation, and anchoring to the initial gaze guidance, allowing segmentation to extend beyond directly fixated regions while limiting semantic drift. GazeRefine requires no segmentation masks, fine-tuning, adapters, prompt encoders, or gradient updates. We evaluate the method on gaze-annotated polyp segmentation and prostate MRI segmentation. The results show strong performance on colonoscopy images and competitive performance on prostate MRI, supporting gaze-guided prototype refinement as a promising approach for segmentation-label-efficient, human-in-the-loop medical image segmentation.
Our tools and code can be found in the following repository \href{https://github.com/MohammedOussamaBEN/GazeRefine/}{\bf{GazeRefine}}

\end{abstract}

\section{Introduction}

Medical image segmentation is essential for computer-aided diagnosis, treatment planning, surgical guidance, and quantitative disease monitoring. Despite substantial progress from U-Net architectures to self-configuring pipelines such as nnU-Net, most high-performing methods still depend on dense pixel-level annotations and task-specific training \cite{ronneberger2015unet,isensee2021nnunet,litjens2017survey,wang2022medicalsegmentation}. This dependence remains a major limitation in medical imaging, where expert masks are costly, time-consuming, subject to inter-observer variability, and difficult to scale across organs, modalities, institutions, and rare pathologies \cite{tajbakhsh2020embracing}. Domain shifts, heterogeneous acquisition protocols, and privacy constraints further limit the repeated collection of annotations and retraining of specialized models for new clinical settings \cite{gibson2018niftynet,karani2020testtime}.

Annotation-efficient segmentation has therefore become an important direction in medical image analysis. Weakly supervised approaches reduce mask requirements through points, scribbles, boxes, or pseudo-labels, but generally still require task-specific training or optimization \cite{bearman2016point,luo2022scribble,cheng2022pointly,tian2021boxinst}. Promptable foundation models such as SAM and MedSAM enable interactive segmentation through explicit visual prompts and segmentation-specific architectures \cite{kirillov2023segment,ma2023medsam}. In parallel, self-supervised vision transformers such as DINO, DINOv2, and DINOv3 provide dense patch representations with \textbf{\textit{emergent localization and semantic grouping properties}} without pixel-level supervision \cite{caron2021emerging,oquab2023dinov2,simeoni2025dinov3}. These representations have motivated training-free segmentation approaches based on frozen features, similarity matching, and feature propagation \cite{sacha2025proto,cohen2022seg,zhang2023pers,liu2024matchersegment,hamilton2022unsupervised,barsellotti2025}. In the medical domain, Gaze2Segment \cite{khosravan2016gaze2segment} demonstrated that gaze-derived cues can support model-free segmentation by combining gaze heatmaps with image saliency. However, many recent interactive approaches still depend on explicit clicks or boxes, segmentation-specific prompt encoders, adapters, fine-tuning, or inference-time optimization \cite{jiang2025promptengineeringsegmentmodel}. 

In this context, clinician gaze provides a natural source of clinical guidance for medical image segmentation. During image interpretation, fixation locations and durations reflect task-relevant visual attention and can indicate suspicious structures or clinically meaningful regions. Recent methods have used gaze either as weak supervision for training segmentation models, as in GazeMedSeg and GradTrack \cite{zhong2024gazemedseg,wang2025gradtrack}, or as an input prompt to SAM-based segmentation models, as in GazeSAM and GazeMedSAMv2 \cite{wang2023gazesamsegment,shmykova2025zeroshotgazebasedvolumetricmedical}. In contrast, we investigate whether clinician gaze can directly steer frozen self-supervised representations without task-specific training or a segmentation-specific architecture.

We introduce \textbf{GazeRefine}, a training-free framework that uses clinician gaze as an inference-time prompt for zero-shot medical image segmentation. GazeRefine converts sparse, duration-weighted fixations into foreground and background priors that initialize semantic prototypes in frozen DINOv3 feature space \cite{simeoni2025dinov3}. These prototypes are iteratively refined through foreground--background discrimination, feature-space affinity propagation, and anchoring to the initial gaze-derived representations. This formulation enables the predicted mask to extend beyond directly fixated regions while limiting drift from the original clinical guidance. The method requires no segmentation masks, task-specific fine-tuning, adapters, prompt encoders, or gradient updates. We evaluate GazeRefine on two publicly available gaze-annotated medical segmentation benchmarks with distinct imaging characteristics: polyp segmentation in colonoscopy images and prostate segmentation in magnetic resonance imaging. The results show that gaze can effectively guide frozen visual representations, particularly when target and background regions are well separated in feature space.

Our contributions are summarized as follows:
\begin{itemize}
    \item We introduce \textbf{GazeRefine}, a training-free and segmentation-label-free framework that uses sparse clinician gaze to guide zero-shot medical image segmentation in frozen DINOv3 feature space.

     \item We combine gaze-derived foreground and background prototypes with recurrent refinement based on foreground--background discrimination and kNN affinity propagation, enabling segmentation beyond directly fixated regions.

\end{itemize}

\section{Proposed Method}

\subsection{Overview}

We introduce \textbf{GazeRefine}, a training-free and segmentation-label-free framework for medical image segmentation. GazeRefine combines clinician gaze with semantic representations extracted from a frozen DINOv3 model~\cite{simeoni2025dinov3} to generate dense segmentation masks without task-specific model training. Given an image \(I\in\mathbb{R}^{3\times H\times W}\) and a set of duration-weighted gaze fixations
\(\mathcal{G}=\{(x_m,y_m,d_m)\}_{m=1}^{M}\), where \((x_m,y_m)\) denotes the fixation location and \(d_m\) its duration, the gaze information is projected onto the DINOv3 patch grid and used to construct foreground and background prototypes in the frozen feature space. These prototypes are iteratively refined through foreground--background discrimination, k-nearest-neighbor (kNN) affinity propagation, and gaze-anchored updates, producing a dense segmentation mask from sparse clinician guidance. The entire pipeline requires no segmentation masks, gradient updates, fine-tuning, adapters, or prompt encoders.

\begin{figure*}[t]
    \centering
    \includegraphics[width=0.9\textwidth]{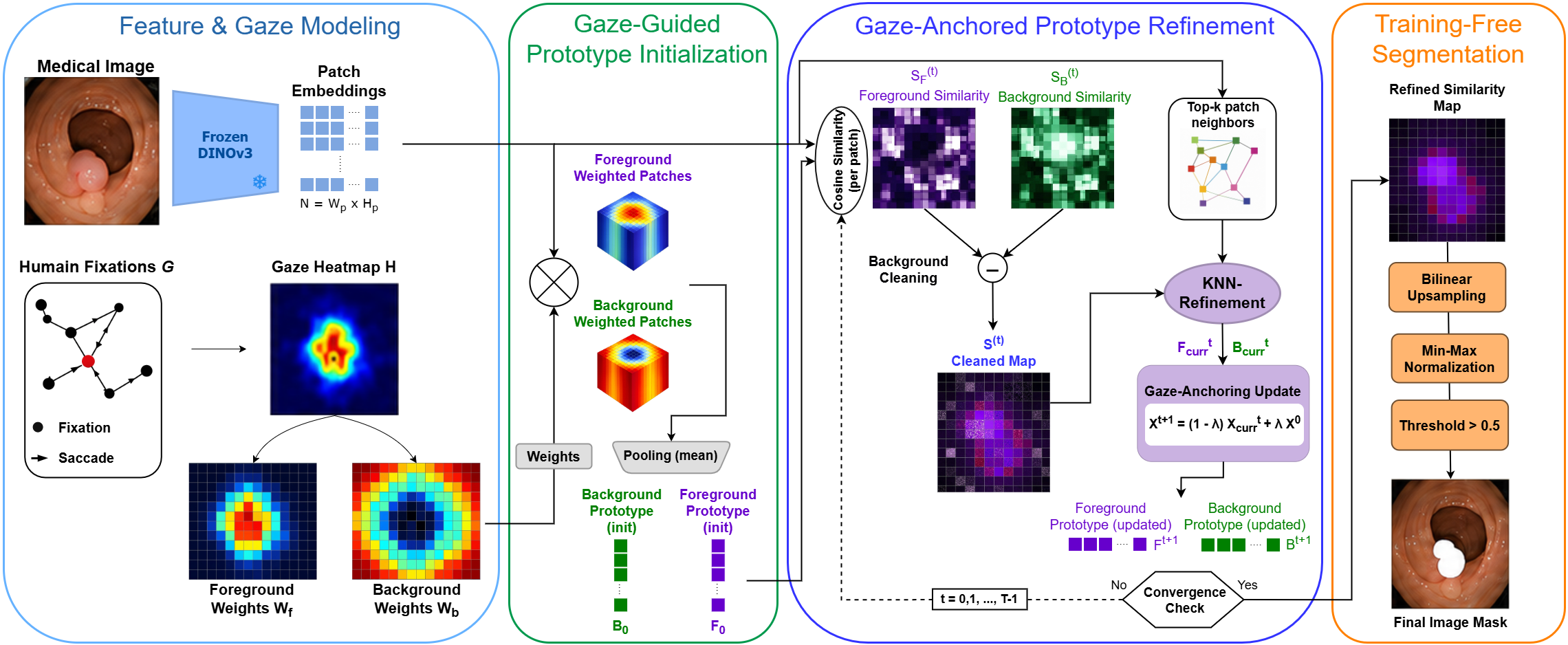}

\caption{Overview of \textbf{GazeRefine}. Clinician gaze and frozen DINOv3 features initialize foreground and background prototypes, which are iteratively refined to generate the final segmentation mask.}
\label{fig:gazerefine_overview}
\end{figure*}

As illustrated in Figure~\ref{fig:gazerefine_overview}, the framework consists of four stages: (i) feature extraction and gaze modeling, (ii) gaze-guided prototype initialization, (iii) recurrent gaze-anchored prototype refinement, and (iv) final mask generation.

\subsection{Feature Representation and Clinician Gaze Modeling}

We map the input image to a pretrained semantic space using a frozen DINOv3 vision transformer~\cite{simeoni2025dinov3}. Let \(P=\{p_i\}_{i=1}^{N}\), with \(p_i\in\mathbb{R}^{D}\), denote the \(\ell_2\)-normalized patch embeddings extracted from the final transformer layer.

In parallel, clinician gaze is converted into a spatial prior. Given the fixation set
\(\mathcal{G}=\{(x_m,y_m,d_m)\}_{m=1}^{M}\), where \((x_m,y_m)\) is the fixation location and \(d_m\) its duration, we construct a duration-weighted gaze map:
\begin{equation}
H(u,v)=\sum_{m=1}^{M}\tilde{d}_m
\exp\!\left(
-\frac{(u-x_m)^2+(v-y_m)^2}{2\sigma^2}
\right),
\end{equation}
where
\(\tilde{d}_m=d_m/(\sum_{r=1}^{M}d_r+\epsilon)\),
\(\sigma\) controls the spatial spread, and \(\epsilon\) ensures numerical stability.

The gaze map is normalized to \([0,1]\) and mapped to the DINOv3 patch grid, yielding \(h_i\in[0,1]\). The initial foreground and complementary background weights are then defined as
\begin{equation}
W^{(0)}_{\mathrm{fg},i}
=
\frac{h_i+\epsilon}{\sum_{j=1}^{N}(h_j+\epsilon)},
\qquad
W^{(0)}_{\mathrm{bg},i}
=
\frac{1-h_i+\epsilon}{\sum_{j=1}^{N}(1-h_j+\epsilon)}.
\end{equation}
These weights initialize the foreground and background prototypes in the frozen feature space.

\subsection{Gaze-Guided Prototype Initialization}

Using the gaze-derived weights, we initialize foreground and background
prototypes through weighted pooling in the frozen DINOv3 feature space. Let
\(\nu(z)=z/(\|z\|_2+\epsilon)\). The initial prototypes are
\begin{equation}
F^{(0)}=\nu\!\left(\sum_{i=1}^{N}
W^{(0)}_{\mathrm{fg},i}p_i\right),
\qquad
B^{(0)}=\nu\!\left(\sum_{i=1}^{N}
W^{(0)}_{\mathrm{bg},i}p_i\right).
\end{equation}
These prototypes serve as gaze-derived foreground and background anchors.

\subsection{Gaze-Anchored Prototype Refinement}
Starting from the initial gaze-derived prototypes \(F^{(0)}\) and \(B^{(0)}\), we refine them for at most \(T\) iterations. At iteration \(t\), the foreground and background similarities of patch \(p_i\) are computed as \(s^{(t)}_{\mathrm{fg},i}=p_i^\top F^{(t)}\) and \(s^{(t)}_{\mathrm{bg},i}=p_i^\top B^{(t)}\), respectively. Since the patch embeddings and prototypes are \(\ell_2\)-normalized, these inner products correspond to cosine similarities in the frozen DINOv3 feature space.

To suppress background responses, we define the non-negative foreground
confidence score
\begin{equation}
S^{(t)}_i=
\max\!\left(
0,\,
s^{(t)}_{\mathrm{fg},i}
- s^{(t)}_{\mathrm{bg},i}
\right).
\end{equation}


The confidence scores are then refined using kNN affinity propagation in the
frozen feature space. We compute the affinities \(A_{ij}=p_i^\top p_j\),
select the \(k\) nearest neighbors \(\mathcal{N}_k(i)\), and obtain
\begin{equation}
\bar{S}^{(t)}_i
=
\sum_{j\in\mathcal{N}_k(i)}
\frac{\exp(A_{ij}/\tau)}
{\sum_{m\in\mathcal{N}_k(i)}\exp(A_{im}/\tau)}
S^{(t)}_j,
\end{equation}
where \(\tau\) controls the affinity distribution among neighboring patches.

The propagated scores are converted into soft foreground confidence weights as \(q^{(t)}_i=\operatorname{sigmoid}(\bar{S}^{(t)}_i)\). The updated
foreground and background weights are
\begin{equation}
W^{(t+1)}_{\mathrm{fg},i}
=
\frac{q^{(t)}_i+\epsilon}
{\sum_{j=1}^{N}(q^{(t)}_j+\epsilon)},
\qquad
W^{(t+1)}_{\mathrm{bg},i}
=
\frac{1-q^{(t)}_i+\epsilon}
{\sum_{j=1}^{N}(1-q^{(t)}_j+\epsilon)}.
\end{equation}

The current prototypes are recomputed as
\begin{equation}
F^{(t)}_{\mathrm{cur}}
=
\nu\!\left(
\sum_{i=1}^{N}
W^{(t+1)}_{\mathrm{fg},i}p_i
\right),
\qquad
B^{(t)}_{\mathrm{cur}}
=
\nu\!\left(
\sum_{i=1}^{N}
W^{(t+1)}_{\mathrm{bg},i}p_i
\right).
\end{equation}

To limit semantic drift, the current prototypes are blended with the initial
gaze-derived anchors:
\begin{equation}
F^{(t+1)}
=
\nu\!\left(
(1-\lambda)F^{(t)}_{\mathrm{cur}}
+\lambda F^{(0)}
\right),
\qquad
B^{(t+1)}
=
\nu\!\left(
(1-\lambda)B^{(t)}_{\mathrm{cur}}
+\lambda B^{(0)}
\right),
\end{equation}
where \(\lambda\in[0,1]\) controls the anchoring strength.

Refinement stops when the maximum number of iterations \(T\) is reached or when both prototypes converge, i.e., when \(\|F^{(t+1)}-F^{(t)}\|_2<\varepsilon\) and \(\|B^{(t+1)}-B^{(t)}\|_2<\varepsilon\). We denote the final iteration by \(T^\star\).
\subsection{Final Segmentation and Training-Free Inference}

After refinement, the score map \(\bar{S}^{(T^\star)}\) is reshaped to the
\(h\times w\) patch grid and upsampled to the original image resolution,
producing a dense response map \(M\). The map is normalized as
\begin{equation}
M_{\mathrm{norm}}
=
\frac{M-M_{\min}}
{M_{\max}-M_{\min}+\epsilon}.
\end{equation}
The final binary mask is obtained by thresholding the normalized response map, i.e., \(\hat{Y}(u,v)=\mathbb{I}\!\left(M_{\mathrm{norm}}(u,v)>\theta\right)\), with \(\theta=0.5\) in all experiments.

\section{Experimental Setup}

\noindent\textbf{Datasets.}
We evaluate GazeRefine on Kvasir-SEG for polyp segmentation~\cite{jha2020kvasirseg} and NCI-ISBI for prostate segmentation~\cite{bloch2015nciisbi}. We follow the gaze annotations and preprocessing protocol of GazeMedSeg~\cite{zhong2024gazemedseg}, which provides fixation data for 1,000 Kvasir-SEG images and 789 NCI-ISBI slices. After removing fixations shorter than 50\,ms and those outside the image boundaries, the sequences contain 5--99 fixations per Kvasir-SEG image and 6--54 per NCI-ISBI slice.

\noindent\textbf{Implementation Details.}
GazeRefine is evaluated in a zero-shot setting with respect to model training: no segmentation labels, parameter updates, task-specific fine-tuning, adapters, or prompt encoders are used. Images are processed at \(592\times592\) resolution using a frozen DINOv3 ViT-L/16 backbone~\cite{simeoni2025dinov3}, producing \(37\times37\) patch embeddings with dimension \(D=1024\). Fixation locations and durations are converted into gaze-derived foreground and background priors for prototype refinement. The same hyperparameters are used for both datasets: \(\lambda=0.5\), \(T=5\), \(k=5\), \(\tau=0.1\), \(\varepsilon=10^{-6}\), and \(\theta=0.5\). The Gaussian spread is set to \(\sigma=2.0\) for Kvasir-SEG and \(\sigma=0.5\) for NCI-ISBI.

\noindent\textbf{Evaluation Protocol.}
For comparison with trainable baselines, we follow the established splits used for each benchmark: 90\%/10\% for Kvasir-SEG and 87\%/13\% for NCI-ISBI~\cite{jha2020kvasirseg,bloch2015nciisbi}. GazeRefine is evaluated only on the corresponding test subsets. Segmentation performance is measured using the Dice  coefficient~\cite{DiceL45} with its standard deviation.
\section{Experimental Results}
\vspace{-3mm}
\subsection{GazeRefine vs. State-of-the-Art Methods}
\vspace{-3mm}
\begin{table}[h!]
\centering

\caption{Comparison with state-of-the-art methods on Kvasir-SEG and NCI-ISBI. Dice (\%).}
\begin{tabular}{lccc}
\toprule
\textbf{Method} & \textbf{Supervision} & \textbf{Kvasir-SEG Dice (\%) $\uparrow$} & \textbf{NCI-ISBI Dice (\%) $\uparrow$} \\
\midrule
U-Net~\cite{ronneberger2015unet} & Full & 82.12 $\pm$ 1.11 & 80.58 $\pm$ 0.48 \\
nnU-Net~\cite{isensee2021nnunet} & Full & \textbf{88.41 $\pm$ 0.47} & 82.43 $\pm$ 0.18 \\
\midrule
PointSup~\cite{cheng2022pointly} & Point & 73.05 $\pm$ 1.64 & 73.46 $\pm$ 4.71 \\
AGMM~\cite{Wu2023SparselyAS} & Point & 75.57 $\pm$ 0.84 & 73.86 $\pm$ 1.26 \\
\midrule
AGMM~\cite{Wu2023SparselyAS} & Scribble & 67.23 $\pm$ 1.02 & 72.70 $\pm$ 1.03 \\
USTM~\cite{LIU2022108341} & Scribble & 66.31 $\pm$ 0.93 & 56.89 $\pm$ 1.23 \\
DMPLS~\cite{luo2022scribble} & Scribble & 69.23 $\pm$ 0.32 & 57.44 $\pm$ 0.56 \\
\midrule
BoxInst~\cite{tian2021boxinst} & Box & 65.72 $\pm$ 2.97 & 73.78 $\pm$ 1.15 \\
BoxTeacher~\cite{cheng2023boxteacherexploringhighqualitypseudo} & Box & 73.33 $\pm$ 1.30 & 75.60 $\pm$ 1.15 \\
\midrule
VAM~\cite{Schneider01061995} & Gaze & 72.21 $\pm$ 0.71 & 78.31 $\pm$ 0.47 \\
VAM\_D-CRF~\cite{krahenbuhl2012efficientinferencefullyconnected} & Gaze & 73.12 $\pm$ 0.60 & 73.86 $\pm$ 1.91 \\
GazeMedSeg~\cite{zhong2024gazemedseg} & Gaze & 77.80 $\pm$ 1.02 & 77.64 $\pm$ 0.57 \\
GradTrack~\cite{wang2025gradtrack} & Gaze & 81.01 $\pm$ 0.66 & 80.25 $\pm$ 0.40 \\
\midrule
GazeSAM~\cite{wang2023gazesamsegment} & \textbf{Zero-Shot} + Gaze & 80.43 ± 0.22 &  82.60 ± 0.17 \\
GazeMedSAMv2~\cite{shmykova2025zeroshotgazebasedvolumetricmedical} & \textbf{Zero-Shot} + Gaze & 85.19 ± 0.13 & \textbf{85.62 ± 0.10} \\
\rowcolor{blue!8}
Ours & \textbf{Zero-Shot} + Gaze & \textbf{88.10 ± 0.14} & 80.28 ± 0.11 \\

\bottomrule
\end{tabular}
\label{tab:kvasir_nci_sota}

\end{table}
Table~\ref{tab:kvasir_nci_sota} compares GazeRefine with fully supervised, weakly supervised, gaze-supervised~\cite{zhong2024gazemedseg,wang2025gradtrack}, and zero-shot gaze-guided methods~\cite{wang2023gazesamsegment,shmykova2025zeroshotgazebasedvolumetricmedical}. All methods are evaluated under the same protocol and on the same test set.
On Kvasir-SEG, GazeRefine achieves a Dice score of \(88.10\% \pm 0.14\), closely approaching the fully supervised nnU-Net result of \(88.41\% \pm 0.47\), obtained using 90\% of the annotated data for training, and outperforming the zero-shot gaze-guided GazeMedSAMv2 result of \(85.19\% \pm 0.13\). GazeRefine also shows lower variability than nnU-Net under the same evaluation protocol.

On NCI-ISBI, GazeRefine achieves a mean Dice score of \(80.28\% \pm 0.11\). Although this is below the best zero-shot gaze-guided result of \(85.62\% \pm 0.10\), it remains competitive with several weakly supervised baselines while relying only on frozen DINOv3 features and gaze-derived priors. The lower performance on prostate MRI may reflect the limited separability of low-contrast anatomical boundaries in general-purpose feature spaces, whereas polyps in colonoscopy images often exhibit more distinctive texture and appearance cues. These results suggest that GazeRefine is most effective when foreground and background regions are well separated in the frozen representation space.
\begin{figure}[t]
    \centering
    \includegraphics[width=0.8\textwidth]{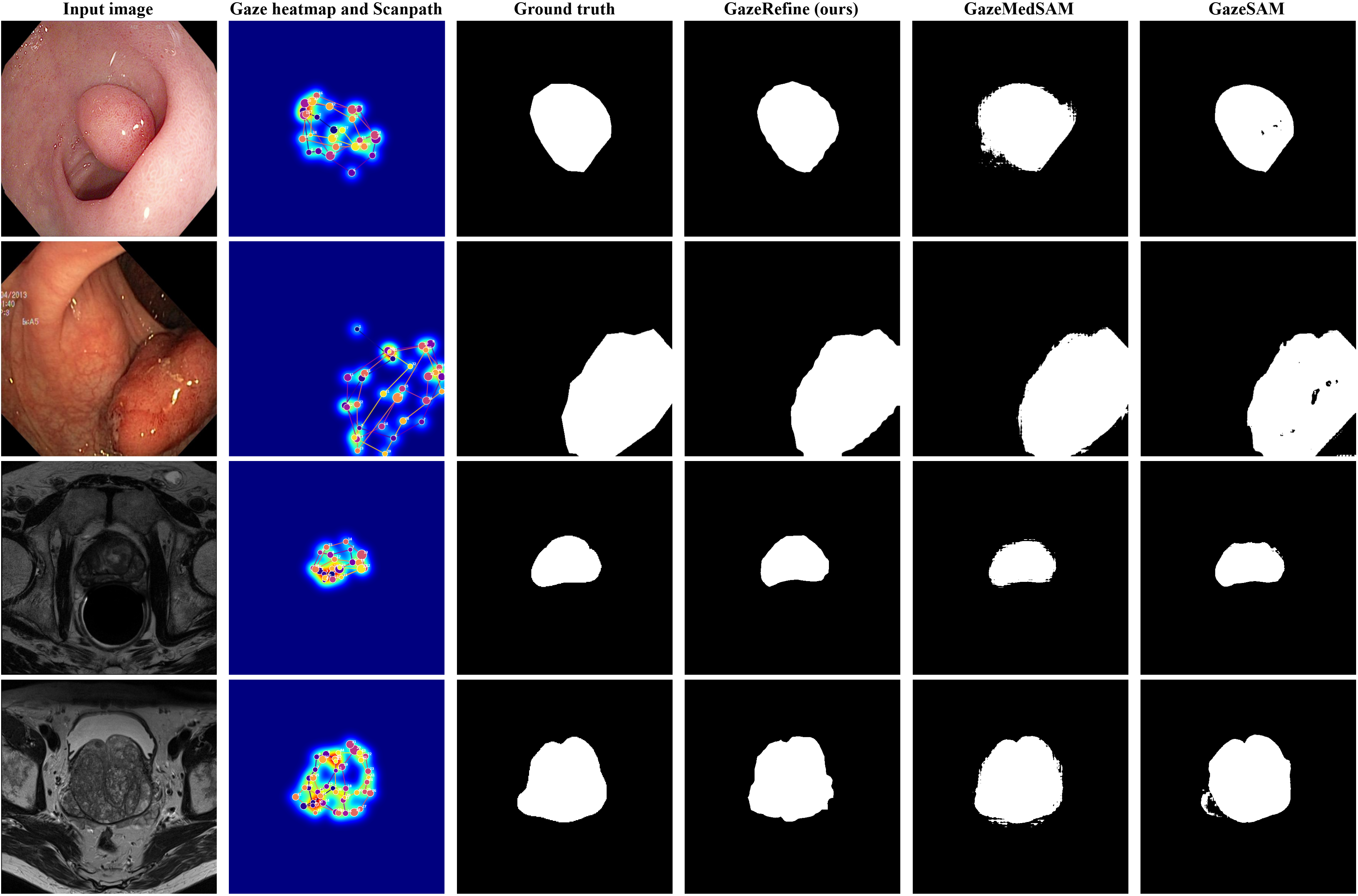}
    \caption{Qualitative comparison of GazeRefine with gaze-guided zero-shot segmentation methods on representative Kvasir-SEG and NCI-ISBI samples.}
    \label{fig:Qualitative_results}
\end{figure}
Figure~\ref{fig:Qualitative_results} presents qualitative segmentation results on representative samples from both Kvasir-SEG and NCI-ISBI. Compared with GazeSAM and GazeMedSAMv2, GazeRefine produces masks that more closely match the ground-truth boundaries while maintaining smoother and more coherent object contours. GazeSAM frequently exhibits under-segmentation and is sensitive to specular highlights and illumination artifacts in endoscopic images. Although GazeMedSAMv2 generally provides better object coverage, both SAM-based methods occasionally introduce spurious foreground regions and boundary artifacts. In contrast, GazeRefine yields cleaner masks with fewer false positives and improved boundary localization. These qualitative results demonstrate that the proposed recurrent prototype refinement effectively enhances foreground-background separation and produces more accurate segmentations across both endoscopic and MRI modalities.

\subsection{Ablation Study} 
\vspace{-2mm}

Table~\ref{tab:ablation} evaluates kNN affinity propagation and background cleaning (\textbf{BG Clean}), following the abbreviations used in the table. The full model achieves \(88.10\%\) Dice on Kvasir-SEG and \(80.28\%\) on NCI-ISBI. Removing kNN propagation reduces performance to \(87.85\%\) and \(73.55\%\), respectively, showing that affinity propagation is particularly beneficial for prostate MRI. Disabling BG Clean causes a much larger drop on both datasets, confirming its importance for limiting background contamination in the prototype space. When both components are removed, performance remains close to the setting without BG Clean, indicating that background cleaning is the dominant refinement component, while kNN propagation provides complementary gains. Gaze is not ablated because it defines the initial foreground and background prototypes and is therefore an intrinsic input to GazeRefine.

\begin{table}[t]
\centering
\caption{Ablation of kNN propagation and background cleaning. Dice (\%).}
\label{tab:ablation}
\footnotesize
\setlength{\tabcolsep}{5pt}
\renewcommand{\arraystretch}{0.95}
\begin{tabular}{@{}cccc@{}}
\toprule
\textbf{kNN} & \textbf{BG Clean} &
\textbf{Kvasir} $\uparrow$ &
\textbf{NCI-ISBI} $\uparrow$ \\
\midrule
\cmark & \cmark & \textbf{88.10} & \textbf{80.28} \\
\cmark & \xmark & 28.70 & 13.50 \\
\xmark & \cmark & 87.85 & 73.55 \\
\xmark & \xmark & 28.35 & 12.79 \\
\bottomrule
\end{tabular}
\vspace{-3mm}
\end{table}

\section{Discussion and Conclusion}
\vspace{-3mm}
GazeRefine performs medical image segmentation directly from frozen DINOv3 patch representations, without relying on SAM-like segmentation models, task-specific prompt encoders, fine-tuning, or gradient updates. Clinician gaze initializes foreground and background prototypes, which are refined through background suppression, affinity propagation, and anchoring to the original gaze guidance.
The results show that clinician gaze can effectively steer frozen visual representations. GazeRefine achieves strong zero-shot performance on Kvasir-SEG, while its lower performance on NCI-ISBI highlights the challenge of applying general-purpose features to low-contrast prostate MRI. This suggests that the method is most effective when foreground and background regions are sufficiently separable in the pretrained feature space.
The ablation study confirms that background cleaning is critical for limiting prototype contamination, while kNN affinity propagation improves spatial coherence. Overall, GazeRefine provides a training-free and segmentation-label-free approach that converts sparse clinician gaze into coherent segmentation masks, supporting clinician-guided prototype refinement as a promising direction for human-in-the-loop medical image segmentation.

\vspace{-3mm}
\section*{Acknowledgment}
\vspace{-3mm}
1- This project was primarily funded by Université Sorbonne Paris Nord.

\noindent
2- This article has been produced with the financial support of the European Union under the LERCO project number CZ.10.03.01/00/22\_003/0000003 via the Operational Programme Just Transition.

\noindent
3- This work was also partially supported by the National Institutes of Health (NIH) under grants R01-HL171376 and U01-CA268808.
\bibliographystyle{splncs04}
\bibliography{references}
\end{document}